# Resonant Quantum Magnetodielectric Effect in Multiferroic Metal-Organic Framework [CH$_3$NH$_3$]Co(HCOO)$_3$

Na Su,[*] Shuang Liu,[*] Yingjie He, Yan Liu, Huixia Fu, Yi-Sheng Chai, and Young Sun[†]

Department of Applied Physics and Center of Quantum Materials and Devices,

Chongqing University, Chongqing 401331, China

[*]These authors contributed equally to this work.

[†]youngsun@cqu.edu.cn

Abstract

We report the observation of both resonant quantum tunneling of magnetization (RQTM) and resonant quantum magnetodielectric (RQMD) effect in the perovskite multiferroic metal-organic framework [CH$_3$NH$_3$]Co(HCOO)$_3$. An intrinsic magnetic phase separation emerges at low temperatures due to hydrogen-bond-modified long range super-exchange interaction, leading to the coexistence of canted antiferromagnetic order and single-ion magnet. Subsequently, a stair-shaped magnetic hysteresis loop along the [101] direction characterizing the RQTM appears below the magnetic blocking temperature. More interestingly, the magnetic field dependence of dielectric permittivity exhibits pronounced negative peaks at the critical fields corresponding to the RQTM, a phenomenon termed the RQMD effect which enables electrical detection of the RQTM. These intriguing properties make the multiferroic metal-organic framework a promising candidate for solid-state quantum computing.

Metal-organic frameworks (MOFs) with an *ABX*$_3$ perovskite-like structure have garnered immense attention over the past decade due to their vast potential in various applications as well as intriguing properties for fundamental science [1-5]. Especially, the variability of organic group *A* and transition metal ion *B* enables the achievement of magnetic and/or electric ordering in perovskite MOFs, offering a new platform for exploring magnetoelectric (ME) multiferroics [5-12]. [(CH$_3$)$_2$NH$_2$]*M*(HCOO)$_3$ (*M* = Mn, Co, Fe, Ni) are the first examples of MOFs exhibiting multiferroicity, where ferroelectric/antiferroelectric ordering develops between 140 and 190 K, and ferromagnetic/antiferromagnetic ordering occurs at much lower temperatures (< 30 K). Because the magnetic and electric orders typically originate from different sites, multiferroic MOFs usually belong to the type-I multiferroics and only weak ME effects were previously observed in them [13-15]. One exception is the multiferroic MOF, [CH$_3$NH$_3$]Co(HCOO)$_3$ (MA-Co), which exhibits magnetically induced electric polarization at low temperatures [12,16], and thus belong to the type-II multiferroic.

In addition to multiferroicity, perovskite MOFs may exhibit unusual magnetic properties such as resonant quantum tunneling of magnetization (RQTM). As discovered in nanomagnets such as ferritin [17], single-molecule [18-21], and single-ion magnets [22-25], RQTM has drawn a lot of interest due to its potential for quantum information and quantum computing [26,27]. Recently, a pronounced magnetodielectric effect was reported in a Dy-based single-ion magnet, which enables the detection of RQTM by an electrical measurement [25]. To date, [(CH$_3$)$_2$NH$_2$]Fe(HCOO)$_3$ (DMA-Fe) has been the only magnetic MOFs showing RQTM [28,29]. In this Letter, we report the discovery of RQTM as well as a pronounced resonant quantum magnetodielectric (RQMD) effect in the type-II multiferroic MA-Co MOF.

Single crystals of MA-Co were prepared by using a hydrothermal method as previously described [30,31]. A stoichiometric solution of Co$^{2+}$/CH$_3$NH$_3^+$/3HCOO$^-$ was prepared by combining three aqueous solutions: CoCl$_2$·6H$_2$O (9 mL, 0.33 M), CH$_3$NH$_3$Cl (9 mL, 0.33 M) and NaHCOO (6 mL 1.5 M). Subsequently, 24 mL of methylformamide (HCONHCH$_3$) was added to provide a suitable environment for crystal growth. The resultant mixture was then transferred into a Teflon-lined autoclave (100 mL) and heated at 140 °C for 72 hours. The resultant pink crystals were filtered from the mother liquor and washed with ethanol. To confirm the structure and phase purity of MA-Co, we conducted powder X-ray diffraction (XRD) at room temperature. The high quality of the single crystal is evidenced by the powder XRD results, shown in Fig. S1 in Supplemental Material [32], which show no secondary phase in the samples. Furthermore, the space group of MA-Co is confirmed as Pnma by the powder XRD pattern.

The MA-Co exhibits a distorted perovskite-like $ABX_3$ structure, as depicted in Fig. 1(a). The metal cations ($B$ = Co$^{2+}$) are interconnected through anti-anti mode bridging HCOO$^-$ anions, forming a distorted B$X_3$ skeleton with three different Co−O bond lengths. The methylammonium CH$_3$NH$_3^+$ cation (MA) is located within the cavities of the distorted [Co(HCOO)$_3$]$^-$ frameworks. Upon cooling to 80 K, the space group changes to $P2_1/c$, and the MA-Co compound undergoes an antiferroelectric-like order due to the order-disorder transition of MA+ accompanied by three fixed hydrogen bonds around the nitrogen atoms [33,34]. As illustrated in Fig. 1(b) and 1(c), there are two types of hydrogen bonds in the cavities: (i) one of the two oxygen atoms of each formate is linked to a nitrogen atom through one linear and stronger N−H···O hydrogen bond within the ac plane; (ii) two oxygen atoms of each formate are linked to the nitrogen atom through two bifurcated and weaker H-bonds along the b-axis [6,18]. Consequently, the influence of hydrogen bonds on the magnetic properties of MA-Co is primarily observed in the ac plane due to the stronger bonds.

Figure 2(a) displays the magnetization along the [101] direction of MA-Co as a function of temperature in both the zero-field-cooling (ZFC) and field-cooling (FC) modes under various magnetic fields. MA-Co exhibits a noticeable magnetic transition at 15.9 K, indicative of antiferromagnetic (AFM) ordering with a weak ferromagnetic component due to spin canting [12]. Additionally, another magnetic transition below 7.5 K is observed in the ZFC curve under 0.05 T. The large discrepancy between the ZFC and FC magnetization indicates a lack of pure long-range magnetic ordering. Similar to the phenomena observed in single-molecule or single-ion magnets, the abrupt decrease in the ZFC magnetization with decreasing temperature in MA-Co is ascribed to a blocking of magnetization. The blocking temperature, denoted by the sudden drop in the ZFC magnetization, shifts to lower temperatures with increasing magnetic field, consistent with typical blocking behavior [20]. Eventually, the blocking behavior disappears in a high enough magnetic field, and the ZFC and FC curves completely overlap.

Figure 2(b) illustrates the dielectric permittivity versus temperature along the [101] direction under various magnetic fields. The dielectric permittivity shows a kink just at the magnetic ordering temperature. Moreover, the dielectric anomaly induced by magnetic ordering becomes more evident in high magnetic fields. The coincidence between magnetic ordering and dielectric permittivity anomaly as well as the clear change of dielectric permittivity under magnetic fields below $T_N$ provide strong evidence for ME coupling in MA-Co.

The temperature dependent magnetic behavior of MA-Co along the [101] direction is similar to that of DMA-Fe where intrinsic magnetic phase separation leads to the simultaneous presence of canted AFM order and isolated single-ion quantum magnets [28]. The isothermal magnetization curve shown in Fig. 3(a) provides further evidence for magnetic phase separation. The *M-H* curve along the [101] direction at 2 K displays clear hysteresis in the low-field region but keeps rising in high magnetic fields. The non-saturating behavior of magnetization is ascribed to the AFM phase. After subtracting the contribution from the AFM component, we obtain a regular stair-shaped hysteresis loop shown in Fig. 3(b), which is a characteristic of RQTM [18,19]. The sharp jumps in magnetization around zero and ±3.5 T field correspond to the occurrence of RQTM.

The magnetic phase separation, i.e., the coexistence of AFM order and isolated single-ion quantum magnets, is due to the selective long-range super-exchange interaction modified by hydrogen bonds, as proposed for DMA-Fe [28]. The $MA^+$ cations in the cavities of the $[Co(HCOO)_3]^-$ framework are linked to three formate ligands through one linear and stronger N-H⋯O hydrogen bond within the ac-plane and two bifurcated and weaker H-bond along the b-axis [33]. The transition metal ions are interconnected through long-range super-exchange paths Co-O-C-O-Co, with the exchange energy dependent on the geometry of the super-exchange path according to the Goodenough-Kanamori rules [35]. Within the ac plane, there are two paths for $Co^{2+}$: one through a pure formate group, linking the $Co^{2+}$ cations by the pure formate group to form a canted antiferromagnetic order via a long-range super-exchange interaction; the other through a formate group coupled with $MA^+$ cations via the stronger hydrogen bond (N-H⋯O) and through Co-O(H)-C-O-Co exchange path. The hydrogen bond will modify the exchange path between $Co^{2+}$ cations: (i) the exchange energy was changed due to Goodenough-Kanamori rules [35]; and (ii) reducing the overlap between $p$ orbitals of oxygen and d orbitals of $Co^{2+}$ as well as between two $p$ orbitals in the middle of the path. As a result, a small proportion of $Co^{2+}$ ions are isolated from the surrounding and behavior like single-ion magnets, similar to the case in DMA-Fe. This characteristic is not observed along the b-axis in the MA-Co MOF because the hydrogen bond along the [010] direction is too weak to significantly affect the exchange path of the $Co^{2+}$ cations and the orbital overlap along the b-axis [12].

The observation of RQTM in magnetic MOFs enables them as a promising candidate for quantum information and quantum computing. However, one critical issue towards quantum computing applications is to find a simple way to transform RQTM into electrical signal. In spintronics, the change in magnetic state is usually transformed into a variation of electrical transport properties. Since magnetic MOFs are good insulators with an extremely high resistivity at low temperatures, electrical transport properties such as the magnetoresistance and the Hall effect are not measurable in them. Therefore, an alternative way is required to detect the change of magnetic state for MOFs.

As MA-CO is a type-II multiferroic, a relatively strong ME coupling could be expected. Previous studies have shown that the sudden jump in magnetization induces a small change in electric polarization [12]. Then, we explored how the dielectric permittivity responses to the magnetization change. As shown in Fig. 2(b), the temperature dependence of dielectric permittivity under magnetic fields already demonstrates clear magnetodielectric effect in MA-Co. Figure 4 shows the magnetodielectric behavior at 2 K and 5 K. For comparison, the *M-H* hysteresis loop and the differential (d*M*/d*H*) curves are also shown in Fig. 4. The sharp peaks in the differential curves correspond to the positions of the RQTM. Just near the critical fields, the dielectric permittivity exhibits significant anomalies. For example, with magnetic field scanning from 6 T to -6 T, the dielectric permittivity shows a kink around 0 T, and a pronounced negative peak around the second critical field (-3.5 T at 2 K and -1.8 T at 5 K). Especially, a sudden jump is observed at both the positive and negative critical fields (±3.5 T) at 2 K. We term this phenomenon as the resonant quantum magnetodielectric (RQMD) effect because the dielectric anomaly is induced by RQTM. Further measurements of the field dependence of dielectric permittivity at 10 K (above the magnetic blocking temperature) revealed no such kind of anomaly as shown in Fig. S3 in Supplemental Material [32].

The concurrence of the dielectric peak and RQTM is ascribed to the intrinsic ME coupling in this type-II multiferroic MOF. Previous studies have shown that the magnetization transition at ~ 3.5 T induces a sharp jump in electric polarization [12], which is attributed to an inverse Dzyaloshinskii−Moriya (DM) effect. At that time, it was not realized that the magnetization transition is actually related to RQTM. The magnetically induced sharp change in electric polarization leads to a significant anomaly in the dielectric permittivity.

In summary, the multiferroic MA-Co MOF exhibits unusual magnetic and dielectric properties. Magnetic blocking and RQTM are observed along the [101] direction, which is ascribed to intrinsic magnetic phase separation consisting of AFM order and single-ion magnets. The selective long-range super-exchange interaction modified by hydrogen bonds plays a pivotal role in the peculiar magnetic properties. More interestingly, the strong ME coupling in MA-Co leads to a striking resonant quantum magnetodielectric effect, characterized by the sharp peaks of dielectric permittivity accompanying the tunneling of magnetization. Thus, RQTM can be electrically detected by measuring dielectric permittivity.

This work was supported by the National Natural Science Foundation of China (Grant No. 12227806), the National Key Research and Development Program of China (Grant No. 2021YFA1400303), and the Fundamental Research Funds for the Central Universities (Grant No. 2023CDJXY-0049).

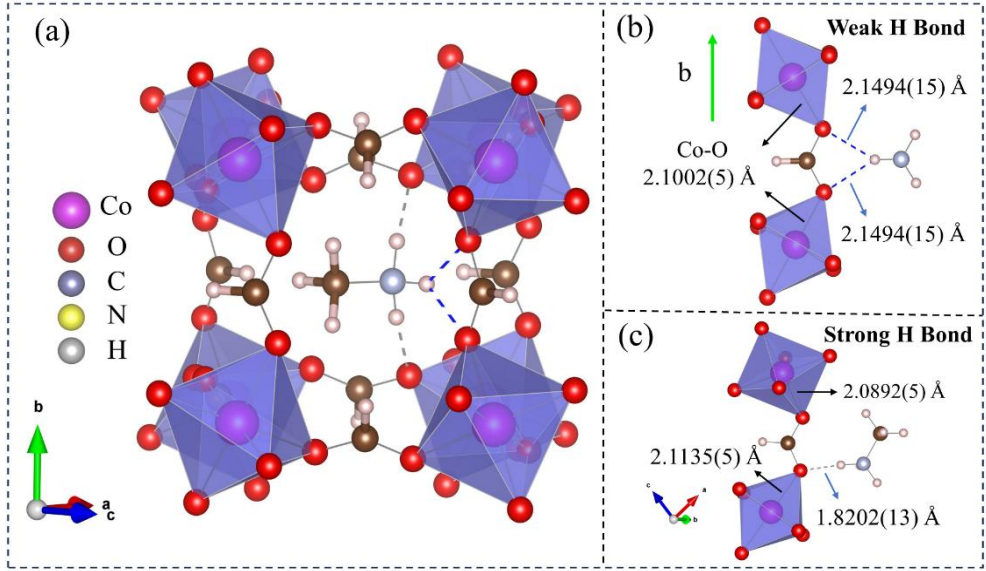

FIG. 1. (a) The ABX$_3$ perovskite-like structure of [CH$_3$NH$_3$][Co(HCOO)$_3$] with MA$^+$ in a metal organic cage. (b) A weak bifurcated H bond along b axis. (c) The strong H bonds in ac plane.

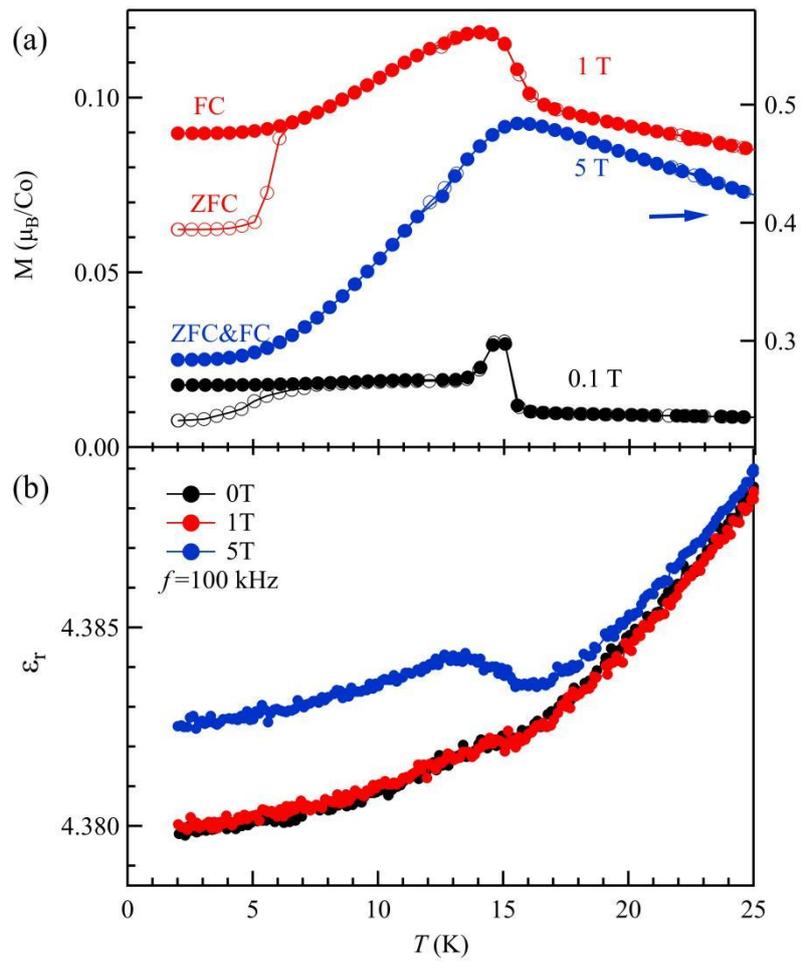

FIG. 2. (a) Magnetization and (b) dielectric permittivity as a function of temperature measured along [101] direction under different magnetic fields.

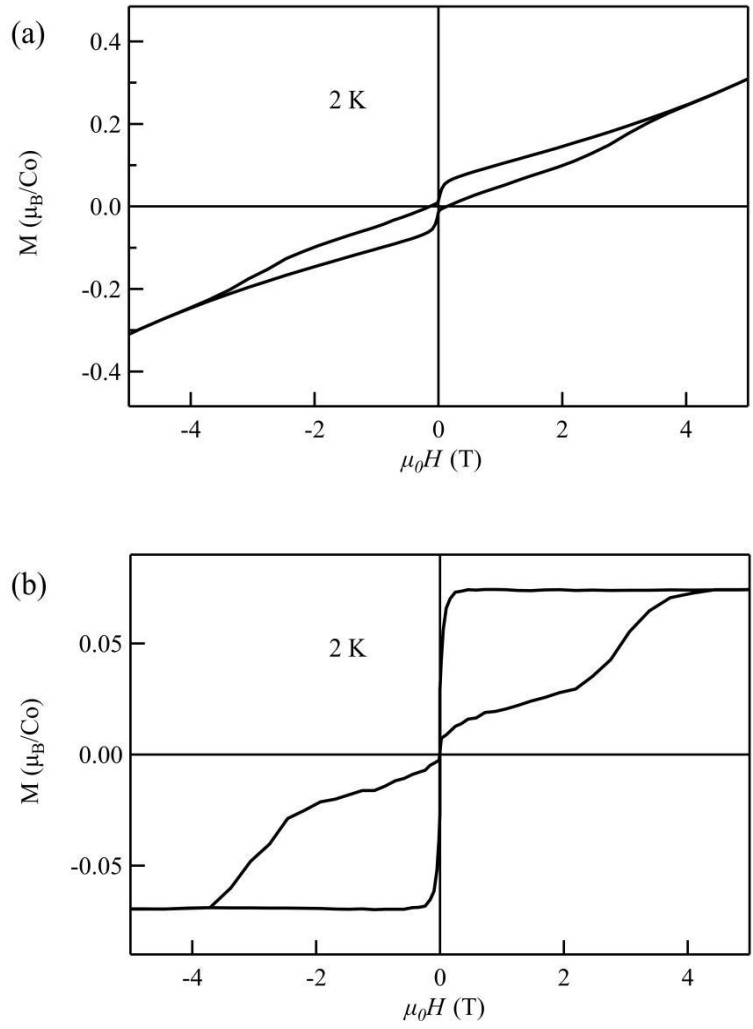

FIG. 3. (a) The *M-H* hysteresis loop along [101] direction at 2 K. (b) The staired-shaped hysteresis loop obtained by subtracting the antiferromagnetic component.

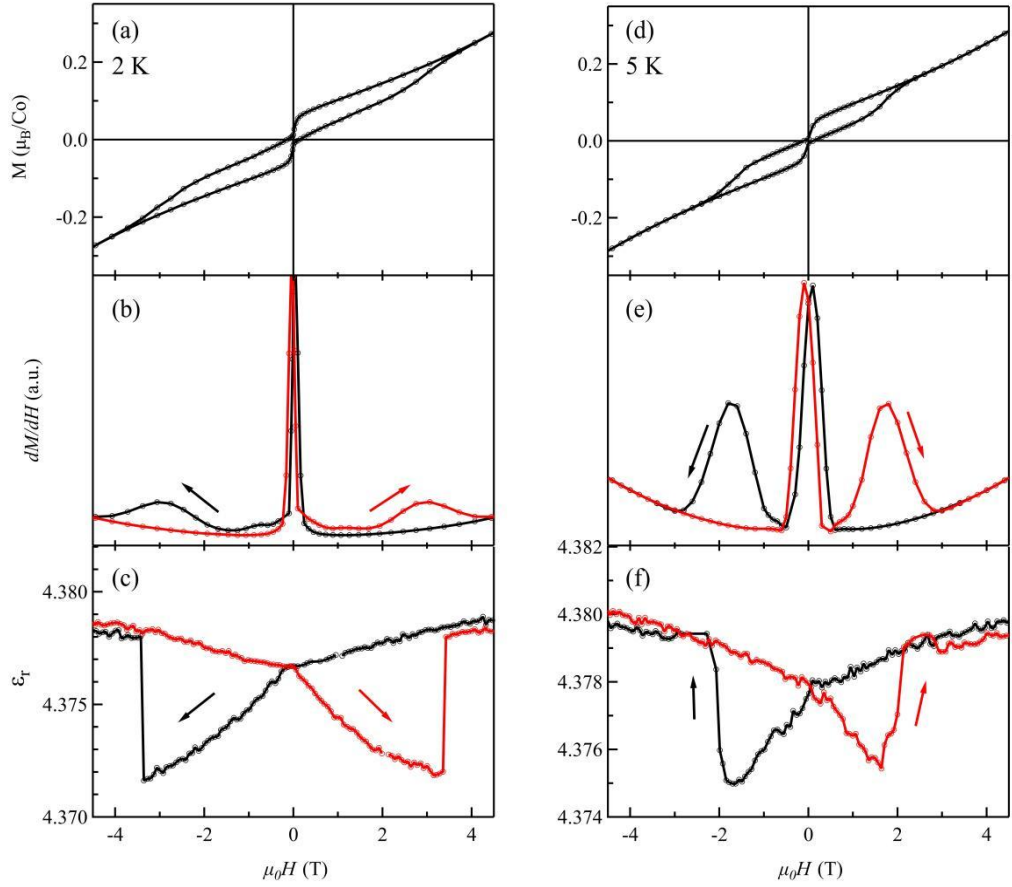

FIG. 4. The *M-H* loop at (a) 2 K and (d) 5 K. The differential of magnetization at (b) 2 K and (e) 5 K. Dielectric permittivity as a function of magnetic field at (c) 2 K and (f) 5 K.